\documentclass[twocolumn,showpacs,preprintnumbers,amsmath,amssymb,prl,superscriptaddress]{revtex4-1}
\usepackage{mathrsfs}
\usepackage{graphicx}
\usepackage{dcolumn}
\usepackage{bm}
\usepackage{amsmath}
\usepackage{amsfonts}

\begin{document}

\title{Topological \emph{p-n} Junction}
\author{Jing Wang}
\affiliation{The Institute of Advanced Study, Tsinghua University, Beijing 100084, China}
\affiliation{Department of Physics, Stanford University, Stanford, California 94305-4045, USA}
\affiliation{State Key Laboratory of Low-Dimensional Quantum Physics, and Department of Physics, Tsinghua University, Beijing 100084, China}
\author{Xi Chen}
\affiliation{State Key Laboratory of Low-Dimensional Quantum Physics, and Department of Physics, Tsinghua University, Beijing 100084, China}
\author{Bang-Fen Zhu}
\affiliation{State Key Laboratory of Low-Dimensional Quantum Physics, and Department of Physics, Tsinghua University, Beijing 100084, China}
\affiliation{The Institute of Advanced Study, Tsinghua University, Beijing 100084, China}
\author{Shou-Cheng Zhang}
\affiliation{Department of Physics, Stanford University, Stanford, California 94305-4045, USA}
\affiliation{The Institute of Advanced Study, Tsinghua University, Beijing 100084, China}

\date{\today}

\begin{abstract}
We consider a junction between surface $p$-type and surface $n$-type on an ideal topological
insulator in which carrier type and density in two adjacent regions are locally
controlled by composition graded doping or electrical gating. Such junction setting on topological insulators
are fundamental for possible device application. A single gapless chiral edge state localized along
the junction interface appears in the presence of an external magnetic field, and it can be probed by scanning tunneling microscopy and transport measurements. We propose to realize this topological \emph{p-n} junction
in (Bi$_{1-x}$Sb$_x$)$_2$Te$_3$, which has insulating bulk properties and a tunable surface state across
the Dirac cone.

\end{abstract}


\maketitle

Topological insulators are new states of quantum matter with a full insulating gap in the bulk and
gapless edge or surface states interesting for condensed matter
physics~\cite{qi2010,moore2010,hasan2010,qi2011}.
The surface states of a three-dimensional (3D) topological insulator are comprised of an odd number
of massless Dirac cones with spin helical structure in the momentum space which are
protected by time-reversal symmetry. Such spin-helical metallic surface states
are expected to host a wide range of exotic quantum phenomena such as Majorana fermions~\cite{fu2008},
image magnetic monopole~\cite{qi2009} and topological magneto-electric effect~\cite{qi2008}.
The single Dirac cone on the Bi$_2$X$_3$ (X = Se and Te) surface~\cite{zhanghj2009,xia2009,chen2009} can be viewed as one quarter of
graphene~\cite{novoselov2005}, and it is predicted to exhibit half-Integer quantum Hall effect~\cite{qi2008},
which is a unique property of a time-reversal symmetry-breaking surface and is determined by the bulk topology.
Extensive efforts such as chemical doping and electric gating have been made to achieve the purely conducting
surface in transport on topological
insulators~\cite{hsieh2009,checkelsky2009,analytis2010,qu2010,chen2010,checkelsky2011,taskin2011,ren2011},
however, they are hindered by intrinsic defects in the materials where Bi$_2$X$_3$ is the significant bulk conduction.

\begin{figure}[b]
\begin{center}
\includegraphics[width=2.9in]{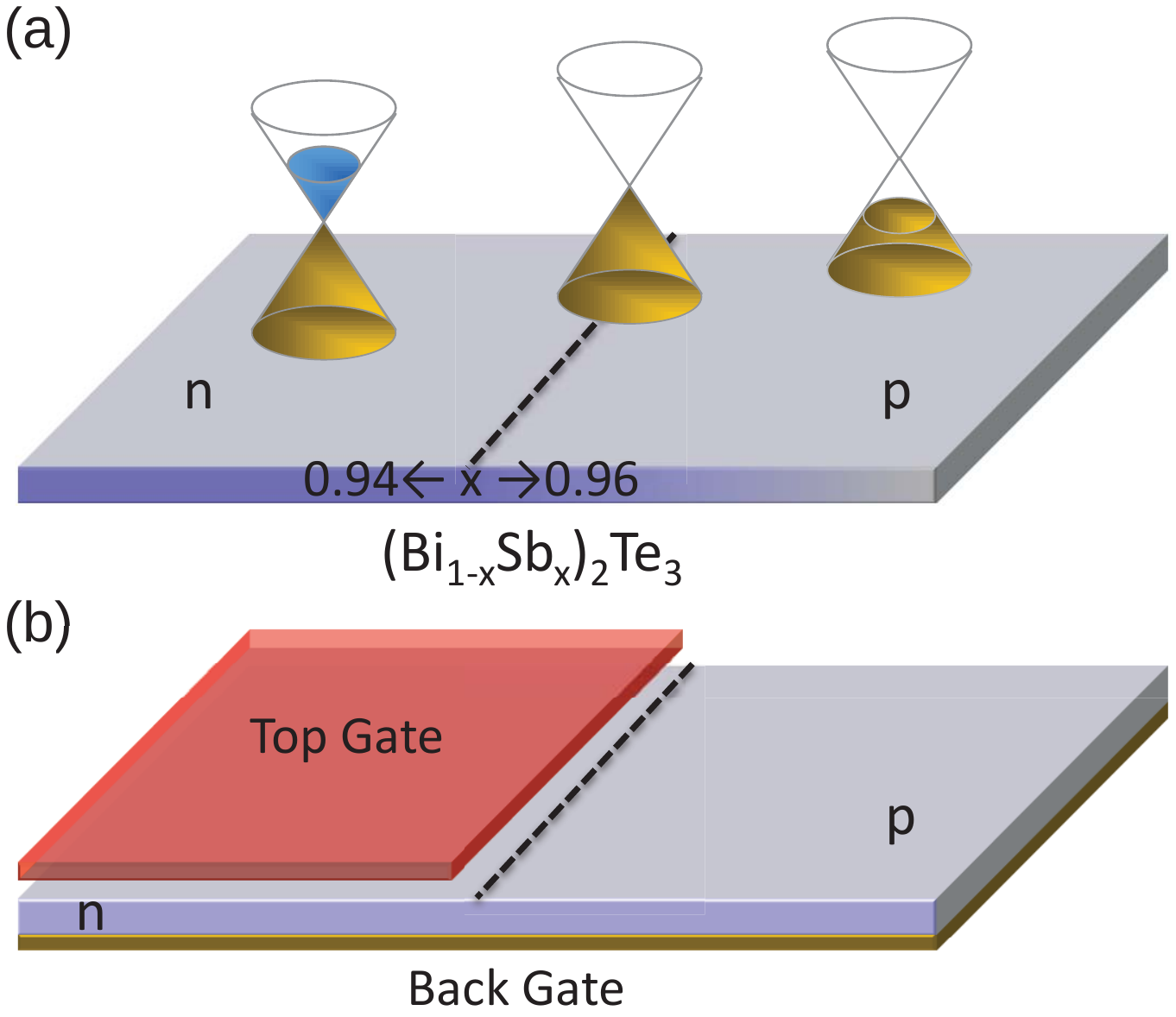}
\end{center}
\caption{(color online). The schematic of the topological \emph{p-n} junction grown by doping of topological insulator alloys
(Bi$_{1-x}$Sb$_x$)$_2$Te$_3$. (a) Compositionally graded doping to achieve spatially variable Dirac cone structure. (b) Electrostatic gating with back gate and top gate to locally control density and carrier type.}
\label{fig1}
\end{figure}

\begin{figure*}[t]
\begin{center}
\includegraphics[width=6.0in]{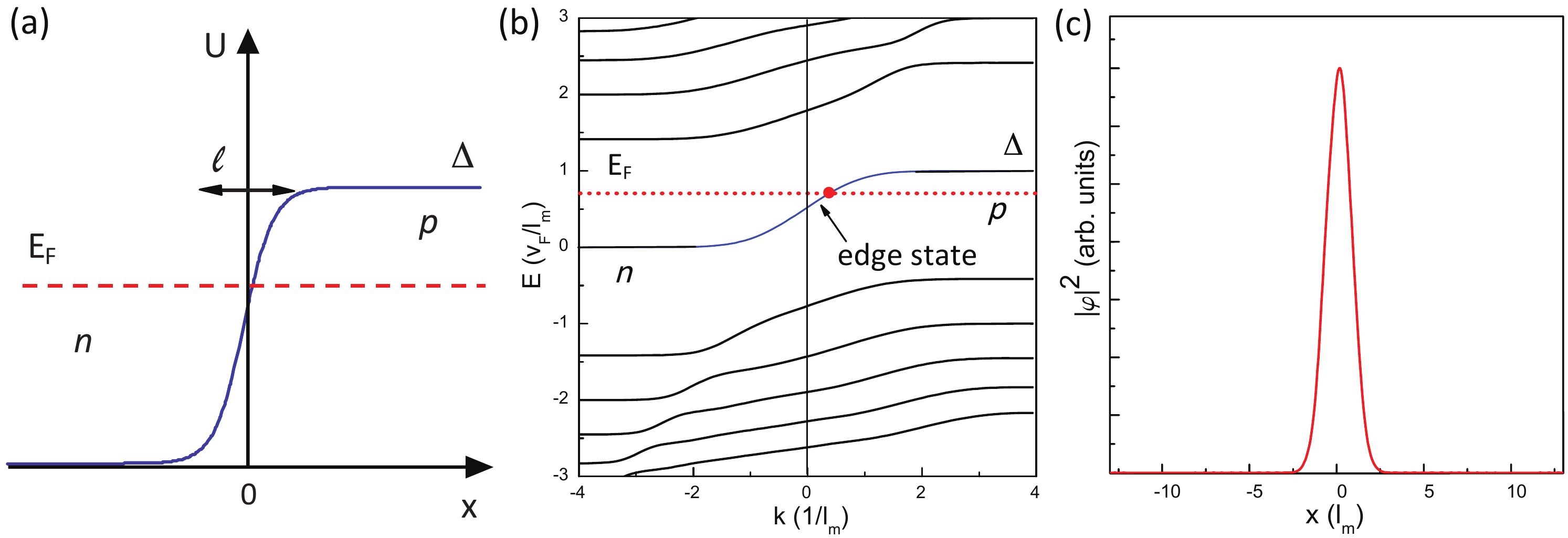}
\end{center}
\caption{(color online). (a) Potential step at the PNJ.
The electrostatic potential $U(x)$ increases from 0 to $\Delta$ over a distance $l$ around $x=0$.
The Fermi level $E_F$ lies in the conduction band in the \emph{n}-doped regime and in the valence
band for the \emph{p}-doped regime.  (c) Energy spectrum at the PNJ. The blue line denotes the gapless chiral edge state confined along the \emph{p-n} interface, where the density of the edge state around $E_F$ is plotted in (d). The flat bands at finite $k$ are just the LLs in the \emph{n} and \emph{p} regions.}
\label{fig2}
\end{figure*}

Recently, three experimental groups have successfully engineered the band structure of topological insulators by
molecular beam epitaxy growth of (Bi$_{1-x}$Sb$_x$)$_2$Te$_3$~\cite{zhang2011,kong2011} ternary compounds and tetradymite Bi$_{2-x}$Sb$_x$Te$_{3-y}$Se$_{y}$~\cite{arakane2012}.
By tuning the ratio of bismuth to antimony, they can get ideal topological insulators with truly
insulating bulk and tunable surface state across the Dirac cone (\emph{p}- to \emph{n}-type) that behave like
1/4 of graphene. By composition graded doping one may form a \emph{p-n} junction~(PNJ) on
a topological insulator surface [see Fig.~\ref{fig1}], which is similar to graphene PNJ~\cite{williams2007,huard2007}.
The properties in such a junction setting of topological insulators are not yet known~\cite{wray2011}.
On the other hand, graphene PNJs are not only promising for transistor,
but also are predicted to host novel phenomena reflecting the massless Dirac character of carriers such
as Klein tunneling~\cite{katsnelson2006} and Veselago lensing~\cite{cheianov2007}.
Therefore, it is important and straightforward to study the physics and relevant applications of PNJ
on a topological insulator surface in analogy of graphene junction. The topological insulator surface
junction is also different from the graphene junction for it only consists of a single Dirac cone.
If a magnetic field is applied perpendicular to the surface junction,
the different types of the carriers would give rise to a quantized Hall conductance $\pm e^2/2h$ in the \emph{n} and \emph{p} regions, respectively, so that a single chiral edge state arises along the \emph{p-n} interface. In the following,
we will refer to such PNJ on a topological insulator surface as `topological \emph{p-n} junction'.
The gapless chiral edge mode localized at the PNJ can be controlled by gating and magnetic field, which may be probed through scanning tunneling microscopy (STM) and transport measurements.

We propose two methods to fabricate the topological \emph{p-n} junction.
One is composition graded doping, the other is electrostatic gating. The compound (Bi$_{1-x}$Sb$_x$)$_2$Te$_3$ with $x=0.94$~(0.96) is an ideal topological insulator with surface \emph{n}~(\emph{p})-doping,
where the surface Dirac cone is in the insulating bulk gap and the Fermi level stays slightly above~(below) the
Dirac cone~\cite{zhang2011}. In analogy to the growth method of PNJ in \emph{p}-GaAs/\emph{n}-Al$_x$Ga$_{1-x}$As,
one can fabricate the junction between \emph{p}-type and \emph{n}-type topological insulators by
composition graded doping from $x=0.94$ to $x=0.96$ to achieve spatially variable Dirac cone structures~[Fig.~\ref{fig1}(a)].
While in electrostatic gating, one can use the global back gate to control
the position of surface Dirac cone in the bulk gap, combined with the local top gate to control density and carrier type
locally [Fig.~\ref{fig1}(b)]. This will result in device configurations with adjacent \emph{n}-type and \emph{p}-type regions on top surface~\cite{statement}, separated by an electrically tunable PNJ, very similar to graphene PNJ~\cite{williams2007}.
Different from the gating in graphene PNJ,
here the capability of the top gate to control the carrier type and density is limited,
therefore it is better to start from (Bi$_{1-x}$Sb$_{x}$)$_2$Te$_3$ with $x=0.96$ rather than Bi$_2$Te$_3$ or Sb$_2$Te$_3$.
The Fermi velocity of Dirac electron in the \emph{p} and \emph{n} regions of such topological \emph{p-n} junction are almost
the same~\cite{zhang2011}. The thickness of depletion layer of the junction at zero bias can be estimated through
Poisson equation $d^2V(x)/dx^2=-{\rho(x)}/{\epsilon}$,
where $\epsilon$ is the dielectric constant, $V(x)$ is the electric potential, $\rho$ is the carrier density.
With the condition that there are no free carriers in the depletion area, one can get the relation between thickness
of depletion layer $l$ and potential barrier $V_d$, $l=\sqrt{2\epsilon(p+n)V_d\delta/enp}$.
$n$~($p$) is the surface electron~(hole) density, $\delta$ is the depth of surface state inside the sample.
With $\delta\sim1$~nm, $\epsilon\sim100$, $eV_d\sim40$~meV, $n\sim p\approx1\times10^{12}$~cm$^{-2}$,
the thickness of the topological \emph{p-n} junction is estimated to be $l\sim9.4$~nm.

The effective model~\cite{zhanghj2009,liu2010} describing the massless 2D Dirac fermion of a
topological \emph{p-n} junction reads
\begin{equation}\label{h}
\mathcal{H} = v_F\left(\hat{\mathbf{z}}\times\boldsymbol{\sigma}\right)\cdot\mathbf{k} + U(x),
\end{equation}
where $v_F$ is the Fermi velocity, $\boldsymbol{\sigma}=(\sigma_1,\sigma_2,\sigma_3)$ are Pauli matrices that
act on the electron spin degrees of freedom, $\mathbf{k}=-i\boldsymbol{\nabla}$ is the canonical momentum
operator on the surface ($x$-$y$ plane), we set $\hbar\equiv1$, $U(x)$ is the electrostatic potential step
at the \emph{p-n} interface, which can be modeled as $U(x)=\left(\tanh(2x/l)+1\right)\Delta/2$, and $\Delta$ is
just the Fermi energy difference between the $p$- and $n$-region~[see Fig.~\ref{fig2}(a)].
On such junction setting, due to the suppressed backscattering on the topological insulator surface,
the interference between the incident and reflected waves will results in a standing wave pattern
on the surface which decays faster than conventional two-dimensional (2D) electron gas, and has been observed by
STM experimentally~\cite{wang2011}.

More interestingly, when a perpendicular magnetic field $\mathbf{B}=B\hat{z}$ is applied to the junction,
a gapless chiral edge state may appear along the interface in the quantum Hall regime. The basic picture of the formation of chiral edge state along the PNJ is easily visualized as skipping orbital semiclassically, the change in sign of charge carriers across the PNJ produces
a change in the direction of the Lorentz force, causing classical trajectories to curve back towards the interface from
both sides. The orbital effect of magnetic field can be obtained
by Peierls substitution $\mathbf{k}\rightarrow\mathbf{k}+e\mathbf{A}$ in Eq.~(\ref{h}),
where $\mathbf{A}=(0,Bx,0)$ is the vector potential. Here we choose the Landau gauge,
thus $\mathbf{A}$ is parallel to the PNJ and vanishes at the interface,
and the canonical momentum $k$ along $y$-axis is a good quantum number.
The energy spectrum for the PNJ is plotted in Fig.~\ref{fig2}(b)
with $l=l_m$ (the magnetic length $l_m\sim\sqrt{\hbar/eB}\sim10$~nm at $11$~T).
The Fermi level intersects with only one chiral channel of mixed electron-hole character with linear
dispersion (blue curve), where the electronlike channel is from the \emph{n} region and holelike channel is from the \emph{p} region, both of them are from the non-degenerate zero mode of Landau levels (LLs),
and contribute to $\pm e^2/2h$ Hall conductance. The density of the chiral edge state is shown in Fig.~\ref{fig2}(c), which is indeed the edge mode confined along the PNJ.
The scale of such chiral edge state is about $10$~nm, which is smooth on the scale of $l_m$.
To see the chiral edge state more clearly we consider a abrupt potential step,
and in order to simplify the notation, we measure energies in units of $\hbar v_F/l_m$ and lengths in units
of $l_m$. Eigenstates of Eq.~(\ref{h}) with magnetic field that decay
for $x\rightarrow\infty$ have the form~\cite{akhmerov2007}
\begin{eqnarray}
\Psi(x,y) &=& e^{iky}\Phi(x+k),\label{wavefunction}\\
\Phi_n(\xi) &=& e^{-(1/2)\xi^2}\binom{\varepsilon H_{\varepsilon^2/2-1}(\xi)}{H_{\varepsilon^2/2}(\xi)},\\
\Phi_p(\xi) &=& e^{-(1/2)\xi^2}\binom{\left(\varepsilon-\Delta\right)H_{(\varepsilon-\Delta)^2/2-1}(\xi)}
{H_{\left(\varepsilon-\Delta\right)^2/2}(\xi)},
\end{eqnarray}
here $H_{\alpha}(x)$ is the Hermite function. The dispersion relation between energy $\varepsilon$
and momentum $k$ follows by substitution of state (\ref{wavefunction}) into the boundary condition
$\Psi_n(0,y)=\Psi_p(0,y)$, and particle current conservation
$\Psi_n^{\dag}(0,y)\sigma_y\Psi_n(0,y)=\Psi_p^{\dag}(0,y)\sigma_y\Psi_p(0,y)$. Thus we obtain
\begin{equation}\label{dispersion}
e^{-\frac{k^2}{2}}
\begin{pmatrix}
\varepsilon H_{\varepsilon^2/2-1}(k)
\\
H_{\varepsilon^2/2}(k)
\end{pmatrix}
=
e^{-\frac{k^2}{2}}
\begin{pmatrix}
\left(\varepsilon-\Delta\right)H_{\left(\varepsilon-\Delta\right)^2/2-1}(k)\\
H_{\left(\varepsilon-\Delta\right)^2/2}(k)
\end{pmatrix}.
\end{equation}
The numerical solution of the Eq.~(\ref{dispersion}) is qualitatively similar to Fig.~\ref{fig2}(c).
For $k\rightarrow-\infty$, the LLs energy in the \emph{n} region are at $\varepsilon_N^2/2=N$, which is
$\varepsilon_N=\pm\sqrt{2}(\hbar v/l_m)\sqrt{N}$; while for $k\rightarrow+\infty$, the energy is
$(\varepsilon_N-\Delta)^2/2=N$, which is $\varepsilon_N=\pm\sqrt{2}(\hbar v/l_m)\sqrt{N}+\Delta$, and the
velocity $v_N=\hbar^{-1}d\varepsilon_N/dk$ vanishes. Those flat bands at finite $k$ in Fig.~\ref{fig2}(c) are LLs in the $n$ and $p$ regions.
The chiral edge state does not exist if the potential barrier is too high,
in that case the \emph{p} and \emph{n} regions are independent,
and the edge channels from both sides can not couple to each other due to energy mismatch. From Eq.~(\ref{dispersion}),
the chiral edge state with $0<\varepsilon<\Delta$ around $k=0$ exists only when $\varepsilon^2/2-1<0$ and
$(\varepsilon-\Delta)^2/2-1<0$, which gives the criterion is
\begin{equation}\label{criterion}
\Delta < 2v_F\sqrt{2eB\hbar},
\end{equation}
where the right part is just the energy difference between the second LL and zero mode. This reflects that increasing the electric field across the junction via gate voltages will destabilize the edge modes. With $\Delta=40$~meV,
the magnetic field $B$ must greater than $2.0$~T in order for such edge state exists.

\begin{figure}[t]
\begin{center}
\includegraphics[width=3.3in]{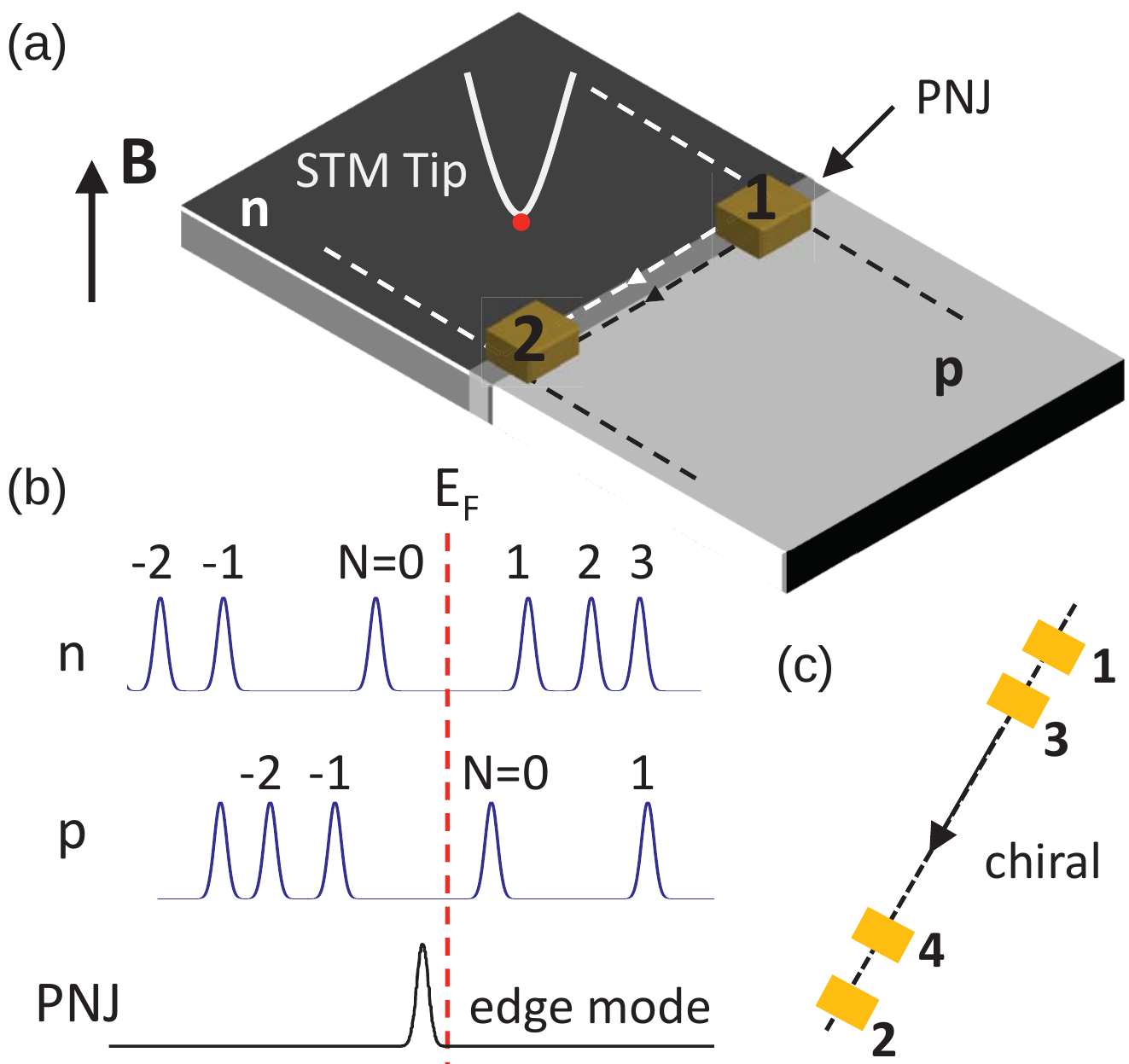}
\end{center}
\caption{(color online). (a) Setup for measuring the gapless chiral edge state along the PNJ.
(b) Schematic of STM measurement, the energy of chiral mode is between the zero mode in the \emph{n} and \emph{p} regions.
(c) Schematic of the chiral mode along \emph{p-n} interface and transport measurement.}
\label{fig3}
\end{figure}

We now propose a practical way to detect such
chiral edge state~[see Fig.~\ref{fig3}(a)]. The state localized at the junction on a topological insulator surface can be imaged by
STM. By applying a magnetic field, the LLs structure of surface states is formed and has already been
observed by STM~\cite{cheng2010,hanaguri2010}.
Discrete LLs appear as a series of peaks in the differential conductance spectrum ($dI/dV$), and the dependence of
the LLs on the magnetic field $B$ shows that the energy of the LLs is proportional to $\mathrm{sgn}(N)\sqrt{\left|N\right|B}$,
where $N$ is the LL index. The energy of zero mode $N=0$ in the \emph{p} and \emph{n} regions can be measured,
between which there are no states in either region as shown in Fig.~\ref{fig3}(b).
Then by moving the STM tip to the PNJ area, one should observe a peak in the $dI/dV$ spectrum with energy
between zero mode in the \emph{p} and \emph{n} regions, and this should be the manifestation of the chiral edge mode.
Moreover, the formation of the chiral edge mode
takes place at a $B>B_{\star}=\Delta^2/8e\hbar v_F^2$, therefore such peak will disappear in the $dI/dV$ spectrum when decreasing the magnetic field.

In addition to the above surface-sensitive technique, one can employ transport measurements. Instead of measuring the
Hall conductance, we can directly measure the longitudinal resistance $R$ along the PNJ. In conventional junction, there is no conducting channel along PNJ. However, in the quantum Hall regime, the single chiral edge mode along topological \emph{p-n} junction will contribute to transport. As shown in Fig.~\ref{fig3}(c), $R$ is measured \emph{along} the PNJ, with current source at 1, ground at 2, and voltage measured between 3 and 4, we will get the longitudinal resistance $R=h/e^2$ contributed from the chiral edge channel. Backscattering is forbidden along this channel, therefore by reversing the current flow at 1 and 2, $R$ will approach infinity. Besides this, when the magnetic field is smaller than the critial value $B_{\star}$, such chiral edge channel will disappear, and we will get zero conductance.

In summary, we propose to use composition graded doping and electric gating in topological insulator alloys
(Bi$_{1-x}$Sb$_x$)$_2$Te$_3$ to fabricate the topological \emph{p-n} junction, where a single 2D Dirac cone junction
on a topological insulator surface may be achieved. A single gapless chiral edge state localized along the \emph{p-n} interface appears in the presence of an external magnetic field, which can be controlled by the gating and magnetic field. The edge mode can be imaged by STM as well as transport measurements. The topological \emph{p-n} devices would lead to novel designs for opto-electronics applications.

We are grateful to X. L. Qi and Yayu Wang for insightful discussion. This work is supported by the
Department of Energy, Office of Basic Energy Sciences, Division of Materials Sciences and
Engineering, under contract DE-AC02-76SF00515. J. Wang is partly supported by the Program of Basic
Research Development of China Grant No.~2011CB921901.

\end{document}